# Large Magneto-Electric Resistance in the Topological Dirac Semimetal α-Sn

Yuejie Zhang,[1,2]* Vijaysankar Kalappattil,[1]* Chuanpu Liu,[1]* Steven S.-L. Zhang,[3]* Jinjun Ding,[1] Uppalaiah Erugu,[4] Jifa Tian,[4] Jinke Tang,[4] and Mingzhong Wu[1]★

[1]Department of Physics, Colorado State University, Fort Collins, Colorado 80523, USA

[2]School of Optical and Electronic Information, Huazhong University of Science and Technology, Wuhan, Hubei 430074, China

[3]Department of Physics, Case Western Reserve University, Cleveland, Ohio, 44106, USA

[4]Department of Physics and Astronomy, University of Wyoming, Laramie, Wyoming 82071, USA

Abstract: The spin-momentum locking of surface states in topological quantum materials can produce a resistance that scales linearly with magnetic and electric fields. Such a bilinear magneto-electric resistance (BMER) effect offers a completely new approach for magnetic storage and magnetic field sensing applications. The effects demonstrated so far, however, are relatively weak or for low temperatures. Strong room-temperature BMER effects have now been found in topological Dirac semimetal α-Sn thin films. The epitaxial α-Sn films were grown by sputtering on silicon substrates. They showed BMER responses that are $10^6$ times larger than previously reported at room temperature and also larger than that previously reported at low temperatures. These results represent a major advance toward realistic BMER applications. The data also made possible the first characterization of the three-dimensional, Fermi-level spin texture of topological surface states in α-Sn.

**Introduction**

Surface electronic states in topological quantum materials exhibit spin-momentum locking. That is, the spin direction of the relevant conduction electrons is locked to the momentum at right angles. In applicable two-dimensional (2D) momentum space, this locking manifests itself as a chiral Fermi contour on which the spin points along the tangential direction everywhere, in either a clockwise or counterclockwise manner, as sketched in Fig. 1(a). Such a Fermi contour usually takes a circular shape. It, however, can turn into a hexagonally warped contour, as in Fig. 1(b), if the material structure has three-fold rotational symmetry.[1,2,3,4,5,6,7] As one walks along such a hexagonal contour, the spin still rotates in a chiral manner as on the circular contour in Fig. 1(a), but now has an out-of-plane component. Such hexagonal warping exists in thin films of topological materials that have (111)-oriented cubic structures or $c$-axis-oriented hexagonal structures.

The spin-momentum locking gives rise to two important effects: (1) prohibition of electron backscattering and (2) highly efficient conversion between charge and spin currents. The first effect opens the possibility of dissipation-free information transport. The second effect allows one to use charge currents to control magnetic properties. Such control is a current subject of intense interest.[8,9,10,11,12,13,14]

There is a third newly reported spin-momentum locking effect: a bilinear magneto-electric resistance (BMER) response.[15,16,17,18] This effect occurs in topological materials with hexagonal warping contours as introduced above. The material exhibits a resistance that (1) scales linearly with the magnetic field, (2) scales linearly with the electric field or current, and (3) varies with the direction of the electric current relative to the crystallographic axes of the material. These kinds of responses were first observed in the topological insulator $Bi_2Se_3$, with a BMER coefficient of $\chi \approx 0.6$ nm$^2$A$^{-1}$Oe$^{-1}$.[15] Here $\chi$ is the scaled BMER response per unit magnetic field and per unit current density. The effect was then also observed for 2D electron gas on surfaces of $SrTiO_3$ crystals, but with a substantially larger coefficient, $\chi \approx 500$ nm$^2$A$^{-1}$Oe$^{-1}$.[17] These observations were made at relatively low temperatures (60 K for $Bi_2Se_3$ and 7 K for $SrTiO_3$). More recent work shows that room-temperature BMER responses are also possible.[18] These later responses were seen in the topological Weyl semimetal $WTe_2$, but were significantly weaker, with $\chi \approx 0.001$ nm$^2$A$^{-1}$Oe$^{-1}$ only.

A basic picture of the BMER effect is as follows:[16] (1) When an electric field **E** is applied, in addition to its first-order correction to the electron distribution that results in a charge current, there is a second-order correction to the distribution that results in equal numbers of electrons populated in the surface states with opposite spins and velocities. This is depicted in Fig. 1(c). The net effect is a pure spin current **J**$_s$ that depends on the square of the electric field, that is, $E^2$. (2) When a magnetic field **H** is applied, the flux of electrons with opposite spins is no longer balanced. As a result, the spin current **J**$_s$ is partially converted to a charge current **J**$_c$, as shown in Figs. 1(d) and 1(e). (3) In Fig. 1(d), **J**$_c$ is along **E** and thereby adds to the ordinary charge current. The net effect is a lower resistance. If **H** is reversed, **J**$_c$ is also reversed. This gives rise to a higher resistance, as shown in Fig. 1(e).

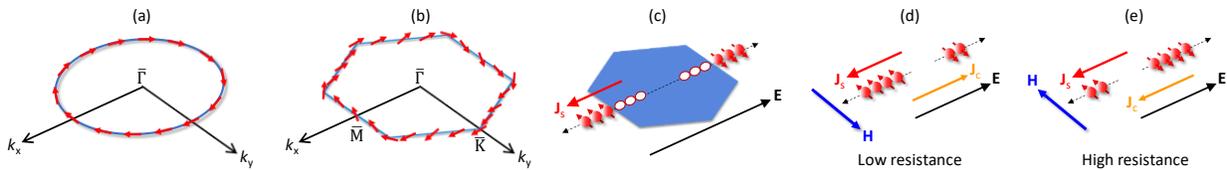

Fig. 1. Bilinear magneto-electric resistance due to spin-momentum locking.[15,16] (a) Spin texture on a circular Fermi contour of topological surface states. (b) Spin texture on a hexagonal warping Fermi contour. (c) Electric field **E**-induced generation of a pure spin current **J**$_s$. (d) Magnetic field **H**-induced partial conversion of **J**$_s$ to a charge current **J**$_c$, giving rise to a lower resistance. (e) In comparison with (d), a flip in **H** leads to a flip in **J**$_c$ and a higher resistance. The short red arrows denote the directions of electron spins.

This article reports a large room-temperature BMER response with a coefficient that is orders of magnitude larger than that in previous work. The experiments used a topological Dirac semimetal (TDS) α-Sn, in contrast to topological insulators, 2D electron gas, and Weyl semimetals in previous experiments. (111)-oriented TDS α-Sn thin films in the 4-6 nm thickness range were grown on single-crystal silicon substrates by sputtering at room temperature. Measurements on these α-Sn films at room temperature showed a resistance that scaled linearly with both the magnetic field and the charge current and strongly depended on the current direction relative to the crystalline axes of the films. The data showed a BMER coefficient of $\chi \approx 2900$ nm$^2$A$^{-1}$Oe$^{-1}$. This coefficient is 10$^6$ times larger than that obtained at room temperature for the Weyl semimetal WTe$_2$.[18] It is also substantially larger than the previous values measured at low temperatures.[15,17] Such a giant effect is most likely due to the low carrier density and spatial asymmetry of the α-Sn films. The data also indicate that in the TDS α-Sn films, the spins on the hexagonal contours lie in-plane at the hexagon vertexes but tilt ±30° out-of-plane at the middle points of the hexagon sides.

Before bringing in the key data that inform the effects outlined above, it is useful to establish three important points. First, the BMER effects offer a totally new approach for information reading and field sensing technologies, but the previous results were weak or only at low temperatures. The very large, room temperature-accessible BMER responses reported here point to the possibility of real applications. This technological significance is further highlighted by the following facts. (a) The effects here have been obtained in a single-material, single-element thin film. This can significantly simplify device fabrication. In stark contrast, other known magnetoresistance (MR) effects, such as giant MR, tunnel MR, and spin-Hall MR, all involve multiple layers of magnetic and non-magnetic materials. (b) The α-Sn films were grown on silicon, a common industrial substrate. (c) The film deposition made use of room-temperature sputtering, an industry-friendly technique. Second, this work shows the technological potential for surface states in TDS α-Sn. There have been major efforts to develop spintronic applications of surface states in topological insulators,[8-14] but corresponding efforts for TDS systems are still largely unexplored. Third, this work also represents the first-ever measurement of three-dimensional (3D) Fermi contours in α-Sn. There have been extensive angle-resolved photoemission spectroscopy (ARPES) measurements on surface and bulk states in α-Sn,[19,20,21,22,23,24,25, 26] but none of them have even touched on the 3D aspect of the Fermi-level spin texture.

**Basic Properties of α-Sn Thin Films**

α-Sn has a diamond cubic crystal structure. Unstrained α-Sn is a gapless semiconductor in which the quadratic conduction and valence bands touch each other at the Γ point near the Fermi level. In the presence of a tensile strain along the [001] or [111] direction, however, the two bands cross each other near the Fermi level. This band crossing forms two Dirac points and give rise to a topological Dirac semimetal (TDS) phase.[22,27,28] One of such Dirac points is shown in Fig. 2. The topological nature of the TDS α-Sn originates from the band inversion: The conduction and valence bands ($\Gamma_8^+$) near the Fermi level are derived from *p* electrons, while the *s* electron-derived band ($\Gamma_7^-$) with opposite parity is below the Fermi level, as illustrated in Fig. 2. The topological surface states (TSS) bridge the $\Gamma_8^+$ conduction band and the $\Gamma_7^-$ valence band,[22,27] as indicated by the red dashed lines.

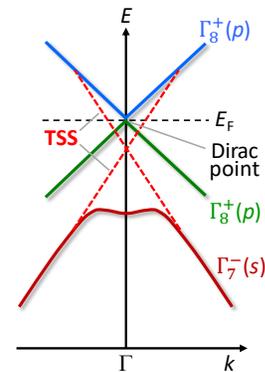

Fig. 2. Schematic of the band structure in topological Dirac semimetal α-Sn thin films.[22,27] "TSS" denotes topological surface states.

In this work, (111)-oriented α-Sn thin films were grown on silicon (Si) substrates by sputtering. The film growth details are given in Section 1 of the Supplemental Materials. The lattice constant of α-Sn (6.489 Å) is larger than that of Si (5.4307 Å). This mismatch yields a perpendicular tensile strain in the α-Sn films. This strain pushes the films into a TDS phase.[22,27,28]

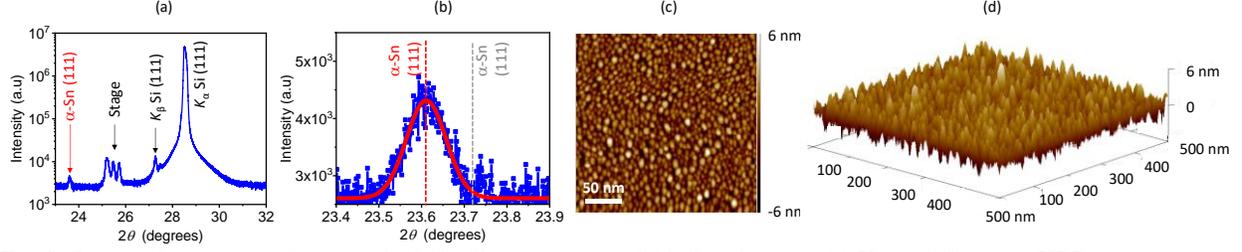

Fig. 3. Structural properties of a 4-nm-thick Sn film grown on a (111) Si substrate. (a) X-ray diffraction (XRD) spectrum. (b) Fitting of the XRD α-Sn (111) peak to a Voigt function. (c) Atomic force microscopy (AFM) surface image. (d) 3D presentation of the AFM image in (c).

Figure 3 presents a quick overview of the structural properties of a representative 4-nm-thick α-Sn film. Panel (a) presents an X-ray diffraction (XRD) spectrum of the film. Panel (b) shows the fitting of the α-Sn (111) peak in (a) to a Voigt function. The red curve is the fit. The red dashed line shows the peak position of the Voigt fit, while the gray dashed line shows the theoretically expected position of the peak. Panels (c) and (d) present 2D and 3D atomic force microscopy (AFM) surface images, respectively.

The data in Fig. 3 establish four important points. (1) The XRD spectrum shows a peak for α-Sn but no β-Sn peaks. This confirms the α phase of the film. As shown in the Supplemental Materials, Sn films with a thickness of 7 nm or larger can exhibit XRD peaks for both the α and β phases, which indicate the coexistence of the two phases. (2) The α-Sn (111) peak is slightly to the left of the theoretically expected peak. This is due to the perpendicular tensile strain in the film, which, in turn, is a consequence of the lattice mismatch at the Sn/Si interface. Analysis yields a tensile strain of 0.46%. It is this strain that makes the α-Sn film a TDS material. (3) The XRD spectrum shows only a single peak for α-Sn. This is a positive indication of epitaxial growth of the α-Sn film. The (111) orientation is critical. As explained above, this produces the hexagonal warping Fermi contour in the film, which is essential to the BMER response. (4) The AFM images demonstrate the granular nature of the film. The grain size is in the 15-20 nm range. The rms surface roughness is about 0.82 nm. Such surface morphological properties are very similar to those of the α-Sn thin films grown by molecular beam epitaxy.[29,30] The significant surface morphology is, in part, responsible for the extremely large BMER response in the film. This is discussed in more detail below.

**Bilinear Magneto-Electric Resistance in α-Sn Thin Films**

Figures 4 and 5 present the main results of this work, namely, the BMER responses obtained on the 4-nm α-Sn film described above. As in previous work,[15,17,18] the experiments were based on Hall bar structures and measurements of the second-harmonic resistance in response to an AC current, as discussed in Section 2 in the Supplemental Materials. The analysis is based on a drive current of the form $I(t) = I_0 \cos(\omega t)$, where $I_0$ is the current amplitude and $\omega$ is the frequency. The total longitudinal resistance is taken to be comprised of an ordinary resistance $R_0$ and a BMER component $CI(t)$. Then, the longitudinal voltage in the α-Sn film can be evaluated as

$$V = [I(t)][R_0 + CI(t)] = [I_0 \cos(\omega t)][R_0 + CI_0 \cos(\omega t)] = \tfrac{1}{2}CI_0^2 + R_0 I_0 \cos(\omega t) + \tfrac{1}{2} C I_0^2 \cos(2\omega t). \quad (1)$$

Division of Eq. (1) by the current amplitude $I_0$ gives the longitudinal resistance as

$$\frac{V}{I_0} = \tfrac{1}{2} C I_0 + R_0 \cos(\omega t) + \tfrac{1}{2} C I_0 \cos(2\omega t). \quad (2)$$

One can see that the coefficient of $\cos(\omega t)$ in Eq. (2) is the ordinary resistance, whereas that of $\cos(2\omega t)$ is one-half the BMER. Simple second-harmonic signal measurements, therefore, allow for a direct measurement of the BMER response. In the following, one uses $R_{2\omega}$ to denote the second-harmonic resistance $\tfrac{1}{2} C I_0$.

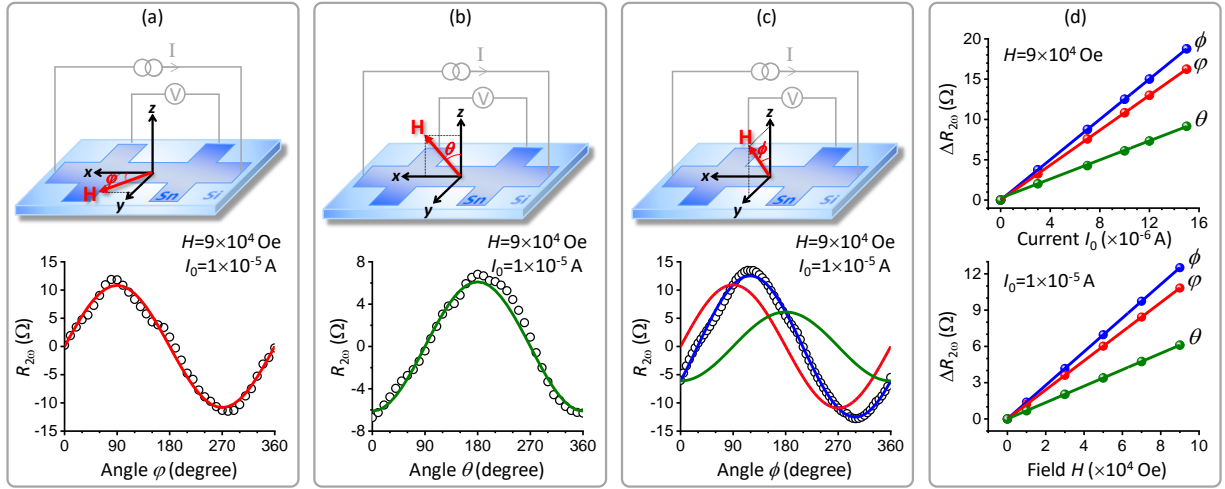

Fig. 4. Second-harmonic resistance ($R_{2\omega}$) as a function of the magnetic field angle for a 4-nm α-Sn film. The circles show the data, the curves show sinusoidal fits, and the lines show linear fits. The field strength $H$ and the current amplitude $I_0$ are indicated. In (c), the red curve shows the fit in (a), the green curve shows the fit in (b); they add together to give the blue curve. In (d), $\Delta R_{2\omega}$ is the amplitude of the sinusoidal $R_{2\omega}$ vs. angle response.

Figure 4 gives $R_{2\omega}$ as a function of the magnetic field angle. Panels (a), (b), and (c) show the $R_{2\omega}$ data measured with rotating the field **H** in three different planes, as indicated. The curves in (a) and (b) are fits to sine and cosine functions, respectively. The red and green curves in (c) are the same as those in (a) and (b). They add together to give the blue curve. Panel (d) shows $\Delta R_{2\omega}$, the amplitude of the sinusoidal $R_{2\omega}$ vs. angle response, as a function of the current amplitude $I_0$ and the field strength $H$.

Five important results are evident from the data in Fig. 4. (1) The data in panel (a) for field rotation in the $xy$ plane indicate that $R_{2\omega}$ takes the maximum and the minimum at $\mathbf{H}||\hat{\mathbf{y}}$ and $\mathbf{H}||(-\hat{\mathbf{y}})$, respectively. This is consistent with the picture depicted in Figs. 1(d) and 1(e). It suggests that the in-plane component of the spin vector points along the $y$ axis. The response in panel (a) also represents a unidirectional magnetoresistance (UMR) along the $y$ axis; the resistance is the largest for $\mathbf{H}||\hat{\mathbf{y}}$ and is the smallest for $\mathbf{H}||(-\hat{\mathbf{y}})$. (2) The data in panel (b) show an UMR along the $z$ axis. This confirms that the spin has a nonzero out-of-plane component. (3) Considered together, the data in Figs. 4(a) and 4(b) suggest that the spin vector lies on the $yz$ plane. This is confirmed by the data in panel (c). Further, the data show a $R_{2\omega}$ maximum at a field angle of $\phi \approx 120°$ and a minimum at $\phi \approx 300°$. This indicates that the spin vector is tilted about 30° out of the $xy$ plane, because the BMER value is the largest for **H** parallel to the spin vector and the smallest for **H** antiparallel to the spin vector. These results are strongly supported by the nice agreement between the experimental response and the blue curve shown in panel (c). (4) $\Delta R_{2\omega}$ scales linearly with both $I_0$ and $H$, as shown in panel (d). (5) The data in panel (b) also exclude the possibility in which an out-of-plane temperature gradient produces $R_{2\omega}$ in the film through the Nernst effect.

Figure 5 compares the $R_{2\omega}$ vs. field angle responses for the currents applied along two different crystalline axes: [11$\bar{2}$] and [1$\bar{1}$0], which correspond to the $\bar{\Gamma}\bar{M}$ and $\bar{\Gamma}\bar{K}$ momentum space lines in Fig. 1(b), respectively. The data in Fig. 5(a) are the same as in Fig. 4, panels (a)-(c). The relevant crystalline axes are defined in Section 1 in the Supplemental Material. The data in Fig. 5(a) show that when the current (or the Hall bar length) is applied along $\bar{\Gamma}\bar{M}$, the spin is in the $yz$ plane and tilts at an angle of about 30° away from the $y$ axis, as discussed above. In contrast, if the current is along $\bar{\Gamma}\bar{K}$, $R_{2\omega}$ exhibits a notable angle dependence for field rotation in the $xy$ plane, but remains almost zero for field rotation in the $xz$ plane, as shown in Fig. 5(b). This indicates that the spin points along the $y$ axis and has a zero out-of-plane component. Such spin orientation is consistent with the spin texture depicted in Fig. 1(b) for a hexagonal Fermi contour.

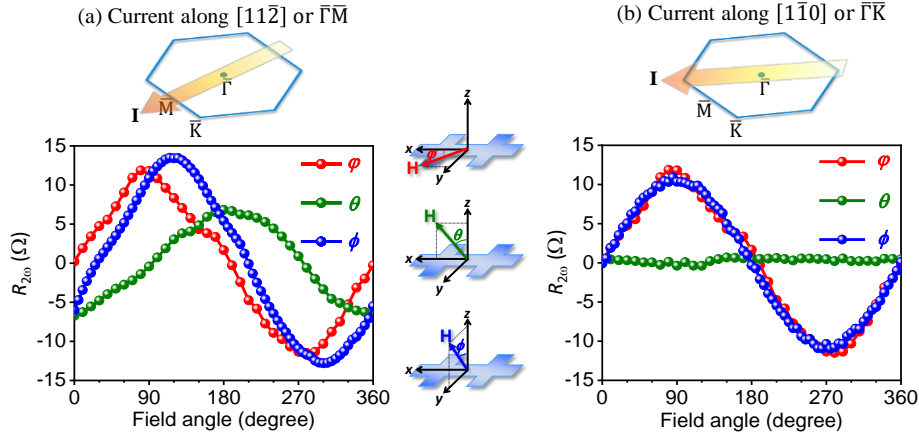

Fig. 5. Second-harmonic resistance ($R_{2\omega}$) as a function of the magnetic field angle for a 4-nm α-Sn film. The data in (a) were measured with a Hall bar whose length is along the [11$\bar{2}$] direction of the α-Sn film, while those in (b) were measured with a Hall bar with its length along the [1$\bar{1}$0] direction.

Taken together, the above results evidently show the presence of the BMER responses in the 4-nm α-Sn film. In fact, the same responses have been observed in multiple samples. As an example, Figure 6 gives the data measured on an α-Sn film that is thicker, with a thickness of 5.8 nm. Figure 6(a) shows the XRD spectrum of the film. It shows an α-Sn (111) peak and shows no peaks for β-Sn. Figure 6(b) shows the $R_{2\omega}$ vs. field angle responses for currents applied along [11$\bar{2}$], which correspond to $\bar{\Gamma}\bar{M}$ in the momentum space. Figure 6(c) gives the data measured for a current direction that was rotated 60° away from [11$\bar{2}$] and was therefore along $\bar{\Gamma}\bar{M}'$ in the momentum space.

The data in Fig. 6 show three key results. (1) The field angle dependences in Fig. 6(b) are about the same as those presented above for the 4-nm film. This consistency supports the results discussed above. (2) After the current direction was rotated from $\bar{\Gamma}\bar{M}$ to $\bar{\Gamma}\bar{M}'$, the $\varphi$ dependence remains the same, but the $\theta$ response flips the sign. This indicates that a rotation in the current direction did not affect the in-plane component of the spin vector, but reversed the sign of the out-of-plane component. (3) $\Delta R_{2\omega}$ is smaller than that for the 4-nm α-Sn film. This result, together with the data in Fig. S3 in the Supplemental Materials, indicates that the BMER decreases with an increase in the film thickness. Possible reasons include that as the thickness increase, the phase of a Sn film transforms from pure α, to a mixture of α and β, and then to β dominant.[31] This is discussed in detail in the Supplemental Materials.

**Bilinear Magneto-Electric Resistance Coefficient**

The strength of the BMER effect can be characterized by a coefficient[17]

$$\chi = \frac{2\Delta R_{2\omega}}{R_0 J_0 H}, \tag{3}$$

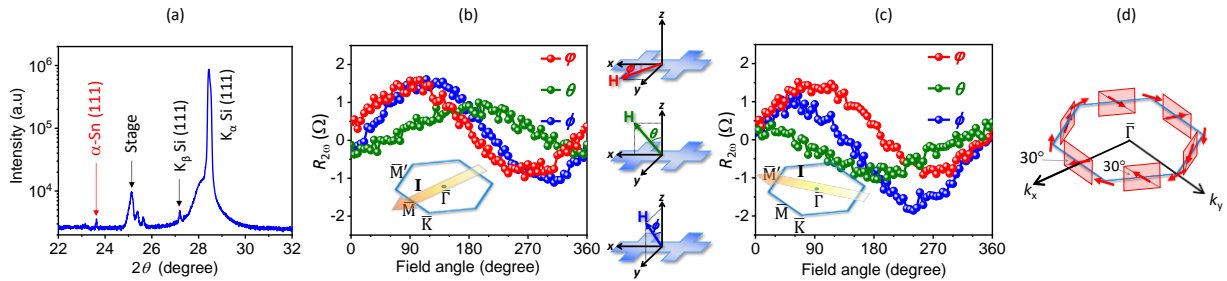

Fig. 6. BMER responses in a 5.8-nm α-Sn film. (a) XRD spectrum. (b) Second-harmonic resistance $R_{2\omega}$ vs. field angle responses measured with currents applied along $\bar{\Gamma}\bar{M}$. (c) $R_{2\omega}$ vs. field angle responses measured with currents applied along $\bar{\Gamma}\bar{M}'$. (d) Spin texture on the Fermi contour.

where $J_0 = \frac{I_0}{wt}$ ($w$ – Hall bar width, $t$ – film thickness) is the current density. With the data in Fig. 4, one obtains $\chi \approx$ 2900 nm$^2$A$^{-1}$Oe$^{-1}$. This coefficient is 10$^6$ times larger than the value measured previously at room temperature for the Weyl semimetal WTe$_2$.[18] It is also markedly larger than the values measured previously at low temperatures.[15,17]

The large BMER response in the TDS α-Sn films possibly originates from the following facts. First, the 2D carrier density ($n$) of the TSS in α-Sn films is relatively low. Previous work has shown that the BMER coefficient $\chi$ increases with a decrease in $n$ in the single relaxation time approximation.[16,17] In fact, a $\chi \propto \frac{1}{n^3}$ response has been observed for the BMER effect in SrTiO$_3$ {see Fig. 4(b) in Ref. [17]}. Previous experiments have shown $n \approx 1.8 \times 10^{12}$ cm$^{-2}$ for TSS in TDS α-Sn films,[31] $3.0 \times 10^{13}$ cm$^{-2}$ for TSS in Bi$_2$Se$_3$ films,[32] and $5.3 \times 10^{13}$ - $7.7 \times 10^{14}$ cm$^{-2}$ for 2D electron gas in SrTiO$_3$.[17] Thus, one can see that $n$ in TDS α-Sn films is more than one order of magnitude smaller than in either Bi$_2$Se$_3$ or SrTiO$_3$.

Second, the α-Sn film in this work does not have spatial inversion symmetry. Specifically, the film bottom surface at the interface with the Si substrate is smooth, while the top surface is relatively rough, as shown by the AFM images in Fig. 3. For this reason, the electrons on the top surface may undergo more frequent scattering from nonmagnetic disorders than those on the bottom surface. The net effect is that the measured BMER responses can be attributed mainly to the spin-momentum locking at the bottom surface. In the case that the film has inversion symmetric, the BMER responses from the top and bottom surfaces would cancel each other and thereby give rise to very weak overall responses, if not zero. This partially explains why $\chi$ in Bi$_2$Se$_3$ films is only 0.6 nm$^2$A$^{-1}$Oe$^{-1}$.[15]

**Spin Texture of Topological Surface States**

Taken together, the data in Figs. 5 and 6 show the 3D aspect of the spins of the Fermi-level TSS in TDS α-Sn. Specifically, the spin in the middle point of a side of the hexagonal Fermi contour is not exactly along the hexagon side, but tilts 30° out-of-plane. As one goes along the hexagonal contour to a vertex, the spin rotates into the plane. As one goes further to the middle point of the next hexagon side, the spin tilts -30° out-of-plane. Such spin rotation is illustrated in Fig. 6(d). There have been considerable ARPES studies on the band structure of α-Sn,[19-26] which include two on spin-resolved ARPES measurements.[19,20] However, none of those studies have touched on the 3D features of the spins on the Fermi contours of the TSS.

The out-of-plane component ($s_z$) of the spin depends on the properties of the TSS at the Fermi level. Such properties include the Fermi velocity ($v_F$), the Fermi wavenumber ($k_F$), and the strength of the hexagonal warping ($\lambda$). If one denotes $\emptyset_k$ as the angle between the $\overline{\Gamma M}$ direction and the wave vector (the current direction), $s_z$ can be evaluated as[1,15]

$$s_z = \frac{\cos(3\emptyset_k)}{\sqrt{[\cos(3\emptyset_k)]^2 + \left(\frac{\hbar v_F}{\lambda k_F^2}\right)^2}}, \quad (4)$$

where $\hbar$ is the reduced Planck constant. The term $\cos(3\emptyset_k)$ dictates the dependence of the BMER on the current direction presented in Figs. 5 and 6, as well as the three-fold symmetry of the chiral spin texture illustrated in Fig. 6(d). For the data in Fig. 5(a) and Fig. 6(b), one has $\emptyset_k = 0$ and $s_z = \sin 30°$. Taken these parameters as well as $v_F = \frac{4}{\hbar} \times 10^{-10}$ eV·m and $k_F = 3.39 \times 10^8$ m$^{-1}$ from previous experiments,[22,31] one can use Eq. (4) to estimate the warping parameter as $\lambda \approx 2$ eV·nm$^3$. This value is about one order of magnitude larger than the values reported for Bi$_2$Se$_3$.[4,6,7,15]

**Final Remarks**

Four remarks related to the above-presented results are as follows. (1) As discussed above, the BMER effect strongly depends on the properties of the TSS at the Fermi level. As a result, future work is of great interest that explores the effects of the Fermi level on the BMER response, with an aim to maximize the effect in the TDS α-Sn. Possible strategies to tune the Fermi level include doping,[19] metallic capping,[24] and voltage gating.

(2) Although this work measured for the first time the spin canting angle and warping parameter of the hexagonal Fermi contour of the TSS in TDS α-Sn, the measurements were somewhat indirect. It is of great interest to directly map the 3D chiral spin texture at the Fermi level in the future.

(3) The hexagonal warping contour concerned in this work is a consequence of the three-fold symmetry of the (111) α-Sn films. In topological insulator α-Sn films that are (001)-oriented and therefore do not have three-fold symmetry, there exist also magnetoresistances that scale linearly with the magnetic and electric fields, as mentioned in Ref. [33]. However, such magnetoresistances have nothing to do with the hexagonal warping, but are associated with electron scattering by spin-orbit structural defects.[33]

(4) The BMER response in this work is also an UMR, as discussed above. However, it is totally different from the UMR effect in bi-layered structures consisting of a magnetic film and a heavy metal or topological insulator thin film.[34,35,36,37,38] They have completely different origins. Further, a recent work reports an UMR in Ge that results from the interaction of an external magnetic field and a pseudo-magnetic field associated with spin-splitted subsurface states in Ge.[39] This UMR scales linearly with both the magnetic and electric fields, but is independent of the current direction relative to the crystalline axes of Ge.

**Acknowledgment**

This work was supported by the U.S. Department of Energy, Office of Science, Basic Energy Sciences (DE-SC0018994). The growth and characterization of α-Sn thin films and the fabrication of the Hall bar structures were also supported by the U.S. National Science Foundation (EFMA-1641989; ECCS-1915849). Work at CWRU was supported by the College of Arts and Sciences at CWRU. Work at UW was supported by the U.S. National Science Foundation (DMR-1710512) and the U.S. Department of Energy, Office of Science, Basic Energy Sciences (DE-SC0020074; DE-SC0021281). The authors thank Carl Patton for valuable suggestions.

*Y.Z, V.K., C.L. and S.S.-L.Z. contributed equally.

★Corresponding author. mwu@colostate.edu

# Supplemental Materials for

# Large Magneto-Electric Resistance in the Topological Dirac Semimetal α-Sn

Yuejie Zhang,[1,2]* Vijaysankar Kalappattil,[1]* Chuanpu Liu,[1]* Steven S.-L. Zhang,[3]* Jinjun Ding,[1] Uppalaiah Erugu,[4] Jifa Tian,[4] Jinke Tang,[4] and Mingzhong Wu[1]★

[1]Department of Physics, Colorado State University, Fort Collins, Colorado 80523, USA
[2]School of Optical and Electronic Information, Huazhong University of Science and Technology, Wuhan, Hubei 430074, China
[3]Department of Physics, Case Western Reserve University, Cleveland, Ohio, 44106, USA
[4]Department of Physics and Astronomy, University of Wyoming, Laramie, Wyoming 82071, USA

## 1. Growth of Sn Thin Films

The Sn thin films were grown on single-crystal (111)-oriented Si substrates by DC magnetron sputtering. The substrates are rinsed sequentially with acetone and isopropyl alcohol before being loaded into the sputtering chamber. Prior to sputtering, the chamber has a base pressure of $2.0 \times 10^{-8}$ Torr; substrate biasing is performed that includes several cycles of Ar ion sputtering of the substrate surface and the post-annealing of the substrate at 250 °C. The Ar ion sputtering process is aimed at removing the thin oxidized layer on the top of the Si substrate, while the post-annealing process is to remove the moisture adherent on the substrate surface. After removing the oxidized layer and the surface moisture, the α-Sn deposition is then carried out at room temperature, at a rate of about 3 nm/min. The sputtering power is set to a moderate value of 10 W, in order to minimize the heating effect during the deposition. The major substrate biasing and sputtering control parameters are summarized in Table S1.

The Si substrates were purchased from MTI Corporation, USA.[1] The substrates are [111] oriented, have a size of 10 mm by 10 mm, and shows a surface roughness smaller than 0.5 mm. The in-plane orientations of the substrates are defined by the notches at the corners of the substrates, as indicated in Fig. S1. These orientations determine the crystalline orientations of the α-Sn films grown on the substrates, because Si and α-Sn both have diamond cubic crystal structures. The Si substrates are *n*-type and undoped, and have an electric resistivity larger than 1 kΩ·cm.

Table S1. Substrate biasing and sputtering control parameters for Sn film growth.

|  |  |  |
|---|---|---|
| Substrate biasing | Ar ion sputtering | 40 W, 2 min |
|  | Annealing | 250 °C, 60 min |
|  | Cycles | 3 |
| Sputtering | Target-to-substrate distance | 6.8 cm |
|  | Sample holder rotation rate | 10 rpm |
|  | Ar pressure | 3 mTorr |
|  | Ar flow | 10 sccm |
|  | Sputtering power | 10 W |

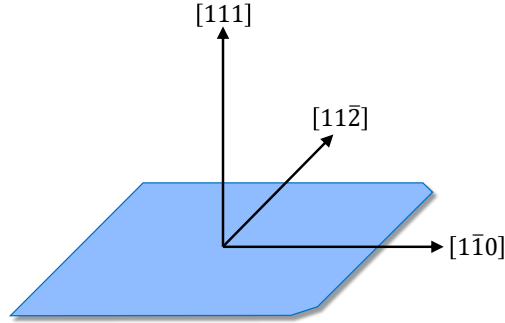

Figure S1. Crystalline axes in the Si substrates used in this work.

## 2. Characterization of Structural Properties of Sn Thin Films

Thin-film characterization has been performed using a Rigaku smart Lab X-ray diffractometer (XRD). The XRD peaks were measured using two different scans. A θ-2θ scan is performed first for a broader range (20° to 80°) to identify the Si (111) peak shift from the theoretical value and adjust the results for instrumental errors. Following that, a 2θ scan (glancing angle XRD) is performed between 20° and 35° to identify α-Sn and/or β-Sn phases. For the glancing angle XRD (GAXRD), the incident angle is kept constant at 0.7°. Thickness calibration for the Sn samples is done through X-ray reflectivity measurements. By fitting the Kiessig fringes obtained from the Fresnel plots using the Rigaku smart lab software, film thicknesses are determined within an accuracy of about 0.1 nm. The thickness values are then used to calibrate the quartz crystal thickness monitor in the sputtering chamber.

The surface morphology of the Sn films is characterized using a Bruker Innova atomic force microscopy (AFM). A surface scan was performed on a 500 × 500 nm surface area, and high-resolution images are taken at 512 × 512 pixels. Height deviation from the mean image data plane is calculated and averaged over about five trails to obtain the root mean square (rms) value of the surface roughness.

## 3. Second-Harmonic Resistance Measurements

In order to measure the electric transport properties of a Sn thin film, a Hall bar device is fabricated through photolithography and argon ion milling processes. Figure S2 illustrates the Hall bar structure. The central area of the Hall bar is 300 μm long and 100 μm wide.

The Hal bar device is connected to a rotatable sample holder and installed in a Quantum Design Dynacool Physical Property Measurement System (PPMS). All electrical contacts are given using indium press-fitting, and all AC second-harmonic resistance measurements are performed using the electrical transport option (ETO) in the PPMS. An AC current of $I(t) = I_0 \cos(\omega t)$, where $I_0$ is the amplitude and $\omega$ = 21.3 Hz is the frequency, is applied to the device, and first- and second-harmonic resistances are measured simultaneously while the magnetic field is rotated in the *xy*, *xz*, and *yz* planes.

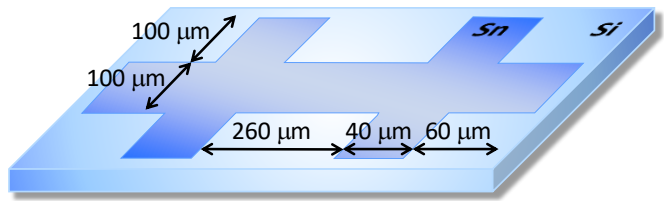

Figure S2. Dimensions of the Sn Hall bar structure.

## 4. Thickness Dependence of Bilinear Magneto-Electric Resistances

Figure S3 compares the bilinear magneto-electric resistance (BMER) responses in four Sn thin films of different thicknesses. Panels (b)-(e) present the second-harmonic resistance $R_{2\omega}$ vs. field angle responses for Sn films of different thicknesses, respectively. The field angles are defined in panel (a). Panels (f) and (g) show $\Delta R_{2\omega}$ vs. current amplitude $I_0$ and field strength $H$, respectively, for the four films. $\Delta R_{2\omega}$ denotes the amplitudes of the $R_{2\omega}$ vs. field angle responses presented in panels (b)-(e). The XRD spectra of the four films are presented in panel (h). The red and green dashed lines in (h) indicate the positions of the α-Sn (111) peak and the β-Sn (100) peak, respectively.

Prior to the discussions of the data shown in Fig. S3, three notes should be made. First, the data in panel (b) are the same as those presented in Fig. 5(a) in the main text. They are repeated to ease the comparison. Second, among the 12 profiles presented in panels (b)-(e), the blue profile in panel (c) shows a distorted-sinusoidal behavior, while the other 11 all show quasi-sinusoidal responses. The abnormal response of the blue profile in panel (c) is probably an instrumental effect. Third, the sputtering growth of the four films and the fabrication of the corresponding Hall bar structures were made under the same conditions at the same time. The 5.8-nm-thick film and the corresponding Hall bar described in the main text were prepared at a different time and are therefore not included in Fig. S3 for comparison.

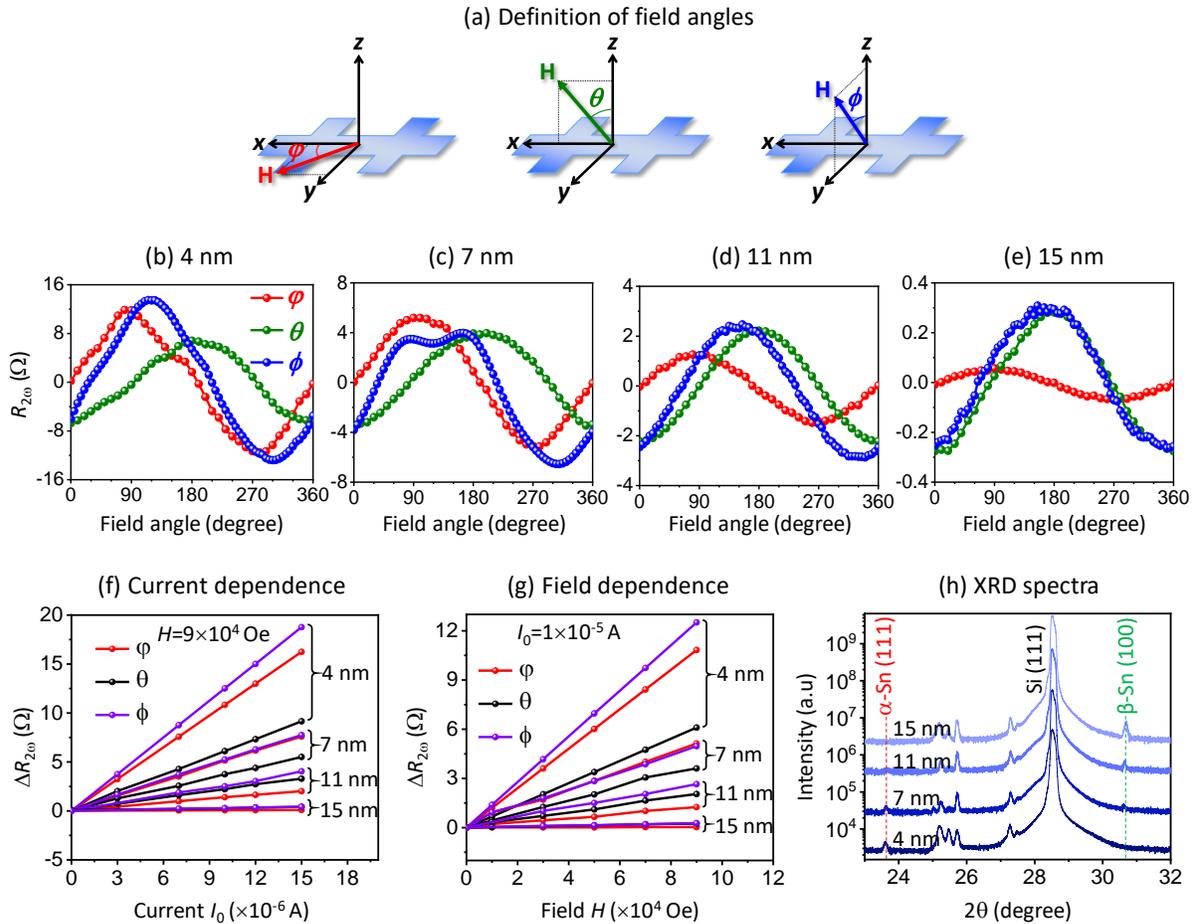

Figure S3. Film thickness dependence of the bilinear magneto-electric resistance effect. (a) Definition of magnetic field angles. (b)-(e) Second-harmonic resistance $R_{2\omega}$ vs. field angle responses measured on Sn thin films of different thicknesses, as indicated. (f) $\Delta R_{2\omega}$ vs. $I_0$ responses for four films of different thicknesses, as indicated. Note that $\Delta R_{2\omega}$ denotes the amplitudes of the $R_{2\omega}$ vs. field angle responses presented in (b)-(e), and $I_0$ denotes the amplitude of the current. (g) $\Delta R_{2\omega}$ vs. field stregnth $H$ responses for the four films. Note that in (f) and (g), the symbols show the experimental data, while the lines are linear fits. (h) The XRD spectra of the four films. The vertical red and green dashed lines indicate the positions of the α-Sn (111) and β-Sn (100) peaks, respectively.

The main result shown by the data in Fig. S3 is that an increase in the Sn film thickness ($t$) gives rise to a notably weaker BMER effect. This is particularly evident in panels (f) and (g) where $\Delta R_{2\omega}$ for the 15-nm-thick film is substantially smaller than that for the 4-nm-thick film. This result is mainly associated with the fact that at room temperature β-Sn is generally more stable than α-Sn, and thin films of α-Sn can be grown on Si substrates because they both have cubic structures.[2] In other words, the realization of α-Sn thin films in this work relies on substrate-enhanced phase stability. For this reason, pure α is present only in ultrathin Sn films, while thicker Sn films can host both the α and β phases, or are β dominant. Such thickness-associated phase stability is clearly seen in panel (h). With an increase in the film thickness ($t$), the α-Sn (111) peak becomes weaker gradually and diminishes at $t = 15$ nm, while the β-Sn (100) peak starts to appear at $t = 7$ nm and then becomes stronger gradually. Since the BMER effect is critically associated with the topological surface states in α-Sn, the gradual transition from α phase to the coexistence of the α and β phases causes a gradual decrease in the BMER strength. Note that the absence of α-Sn peaks in the XRD spectrum of the 15-nm film does not necessarily mean that the film has pure β phase and α phase is completely absent, but may mean that the β phase is dominant over the α phase. Future work to check this is of great interest.

Two additional results are evident in Fig. S3. First, in spite of the weakening of the BMER effect due to the thickness increase, the BMER scales linearly with both the magnetic field and the electric current for all the four samples. This indicates that the responses in the thicker films, though relatively weak, also result from the BMER effect, rather than other magnetoresistance effects. Second, the data suggest that the ratio of the *y* component of the spin vector at the Fermi level to the *z* component decreases with an increase in *t*. The actual reason for this observation is unknown currently, but it is not unreasonable if one considers that the spin canting strongly depends on the properties of the topological surface states at the Fermi level, while those properties are expected to vary if the film undergoes a phase transition from α to a mixture of α and β as well as a change in the strength of the tensile strain in α-Sn. Here the Fermi properties include the Fermi velocity, the Fermi wave number, and the warping strength of the hexagonal Fermi contour, as shown in Eq. (4) in the main text.

*Y.Z, V.K., C.L. and S.S.-L.Z. contributed equally.

★Corresponding author. mwu@colostate.edu